\documentclass{iaus}
\usepackage{graphicx}

\title[Star Formation in Interacting Galaxies] 
{Star Formation in Mergers and Interacting Galaxies:
Gathering the Fuel}

\author[Struck]   
{Curtis Struck$^1$}

\affiliation{$^1$Department of Physics and Astronomy, Iowa State University,
Ames, IA 50011, USA \break email:  curt@iastate.edu}

\pubyear{2006}
\volume{237}  
\pagerange{119--126}
\date{?? and in revised form ??}
\jname{Triggered Star Formation in a Turbulent ISM}
\editors{B. G. Elmegreen \& J. Palous, eds.}
\begin{document}

\maketitle

\begin{abstract}
Selected results from recent studies of star formation in galaxies at different stages of interaction are reviewed. Recent results from the Spitzer Space Telescope are highlighted. Ideas on how large-scale driving of star formation in interacting galaxies might mesh with our understanding of star formation in isolated galaxies and small scale mechanisms within galaxies are considered. In particular, there is evidence that on small scales star formation is determined by the same thermal and turbulent processes in cool compressed clouds as in isolated galaxies. If so, this affirms the notion that the primary role of large-scale dynamics is to gather and compress the gas fuel. In gas-rich interactions this is generally done with increasing efficiency through the merger process.
\end{abstract}

\firstsection 
\section{Introduction}

Star formation (SF) in interacting galaxies (IGs) occurs in a vast range of environments, from the dense, turbulent nuclei in major merger remnants to relatively diffuse regions in (literally) far flung tidal tails. In a short paper, it is impossible to review the detailed processes that orchestrate star formation in these different circumstances. Rather, it seems better to try to give an overview of the "big picture," while keeping in mind the key question - what if anything is different about SF in IGs relative to isolated galaxies?

Even with this restriction it is still helpful to break this large topic up into smaller subtopics, and this is easy to do in two parameter dimensions. The first concerns the distribution of the star formation, i.e., it is natural to consider compact SF in galaxy cores separately from extended SF, which occurs primarily in waves. The second classification dimension is the relative time or merger stage of the interaction, which is possible because most significant interactions lead to merger after one or two close encounters.  Here it is sufficient to distinguish: 1) early stage encounters, from the onset of the interaction up to the second close approach, 2) intermediate stage encounters, up to the time when the bodies of the two galaxies no longer separate, and 3) late encounter stage, which includes the final merger and continuing relaxation. 

\section{Compact Induced Bursts}

Compact SF, including nuclear starbursts, are the source of much of the emission in most LIRGs and ULIRGs, which are the most spectacular examples of interaction induced SF. There is, of course, a huge literature on these objects, and on the topic of ULIRGs I would refer the reader to the recent review of \cite{lon06}. Here I would merely remind the reader that there is now a great deal of evidence confirming that the ULIRG phenomenon generally occurs at the intermediate to late stages of a major merger between two galaxies which both contain a gas-rich disks (very ``wet'' mergers in the current jargon). More specifically, the phenomenon is usually the result of a super-starburst (with perhaps an AGN contribution), triggered by the direct interaction of the two disks in the core of the developing merger remnant (see e.g., the recent models of \cite{bek06}, \cite{hop06}). We can expect that this environment is characterized by strong shocks and extreme turbulence as a result of high velocity intersecting flows. 

This extreme environment has been historically very difficult to observe because it is compact and deeply buried in dust. The latter fact means that we must observe at long wavelengths (mid-IR to radio), and until the launching of the Spitzer Space Telescope there have been few long-wavelength instruments with the necessary sensitivity and resolution. Similarly, with a huge range of spatial scales involved in the dynamics and a plethora of heating, cooling, and feedback processes, it is very difficult to model super-starbursts, except in a partial or schematic way. Detailed modeling of individual systems, like Arp 220, is not yet feasible. 

At the same time, we have increasing numbers of observational clues. Among these is the development of starburst activity with merger age along the Toomre sequence demonstrated by \cite{cha01} using ISO data (see Figure 1). This result highlights the relation between the large scale dynamics and the SF activity. On the other hand, the  \cite{ken98} result that the SFR in ULIRGs follows the Schmidt Law dependence on gas surface density suggests that the fundamental physics of these bursts does not differ greatly from SF in isolated late-type disks despite their violent hydrodynamics. In the coming years the Herschel mission should help us understand this physics.

\begin{figure}
 \includegraphics[bb=-3.0cm 0.0cm 3.0cm 6.0cm]{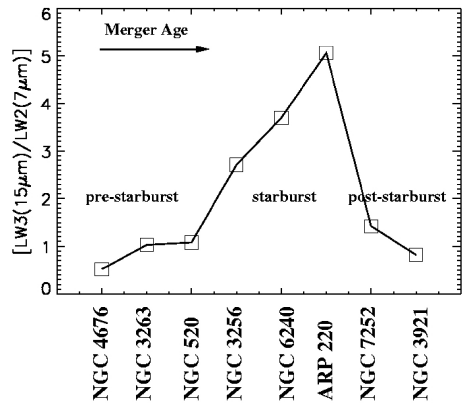}
  \caption{An ISO indicator of star formation rates along the Toomre sequence from \cite{cha01}.
  \protect\\}
\end{figure}

The mysteries of LIRGs are no fewer, though their infrared luminosities are less. For one thing, members of the LIRG class seem to be very heterogeneous. Recent surveys suggest that a significant fraction are not involved in interactions, though probably the majority are. Among the latter, we can understand some of the variety just by assuming that ULIRGs generally pass through a LIRG stage in building up to their full strength, and another in decaying after reaching their peak. The duration of the former could be understood as the result of bulk compression of the galaxies from the time that the central parts of their halos overlap, or the beginnings of disk-disk interaction. The latter phase may be shortened by prompt feedback effects, but much work is needed to confirm such speculations. 

In addition, there are also examples of early stage interactions with core starbursts that reach LIRG levels of intensity. NGC 7469 provides a ring galaxy example, though an active nucleus also contributes in that case (see \cite{wee05}). 

Finally we note that \cite{hin06} have recently pointed out that some LIRGs seem to have SF efficiencies comparable to ULIRGs, but they occur in smaller (less massive) galaxies. Thus, some nearby LIRGs may simply be down-sized ULIRGs. 

\section{Extended Star Formation in Early Stage Interactions}

At the opposite end of the spectrum of SF activity is the question of how interaction induced SF begins, or alternately, how much is SF enhanced at the early stages of interaction? There has been a good bit of work on these questions recently, and some debate on the answers. I have reviewed this discussion recently in more detail than space allows here, so I will merely note a few highlights, and refer the reader to \cite{str06} for more.  Several recent studies of SF in galaxy groups, based on the Sloan or 2dF surveys, have many thousands of objects, and thus, their conclusions are statistically very significant. They reveal modest SF enhancements, especially due to core starbursts, which increase with galaxy-galaxy proximity, and possibly inversely with relative velocity (e.g., \cite{nik04}). The bad news is that these studies may be answering somewhat different questions than those posed above.

For example, many of the galaxies in these groups are not (obviously) interacting, so the proximity effect may be diluted by (possibly nonrandom) projection effects. More seriously, they include interactions at many different stages. Indications of simultaneous proximity and velocity effects suggest systems consummating their merger at intermediate to late stages. 

The GO Cycle 1 Spitzer Space Telescope project "Spirals, Bridges and Tails" (SB\&T), that my collaborators and I have been working on, takes a different approach to these questions. In this project, we have been attempting to use Spitzer imagery and spectra to study the sites and modes of induced SF in an Arp Atlas sample of quite strongly interacting, but pre-merger galaxies. We have a sample of about three dozen systems, and make comparisons to a control sample of isolated disk galaxies drawn from the SINGS legacy survey. The object identifications and images can be viewed at the website - http://www.etsu.edu/physics/bsmith/research/sbt.html. Complementary GALEX and ground-based H$\alpha$ observations of a similarly sized and overlapping sample are underway.

In this study we find clear and statistically significant differences in the 3.6$\mu$m - 24$\mu$m and 8$\mu$m - 24$\mu$m colors of the SB\&T systems and non-interacting spirals.  The distribution of interacting components has a significant red tail in this color, which in these bands suggests more emission from dust heated by embedded young stellar populations. (For more details see \cite{smi06}.) This result is supported by the fact that the IRAS fluxes of the interacting sample are about a factor of 2 larger (per galaxy) than those of the comparison sample, in agreement with the results of \cite{bus88} on a similar sample.

Many of the systems with the reddest [8] - [24] colors appear to be core starbursts in 24$\mu$m images, with disturbed disks or mass transfer bridges that may indicate gas transferred to core regions. Thus, the enhancements we see appear to confirm the importance of dynamical triggering from the time of the first close pass. We do not see any evidence of a proximity effect in this relatively small and specialized sample. The ongoing SF found in tidal features is qualitatively similar to that found in disks, confirming the results of \cite{sch90}. Less than 10\% of the net SF is found in tidal features. 

Every wave formed in a galaxy collision is a unique laboratory of SF processes, and much can be learned from detailed study of individual interacting systems. Given the limited space, I would like to consider one SB\&T system in a bit more detail, the Arp 82 system. This system has been studied in detail by SB\&T collaborator Mark Hancock (see \cite{han06}). 

The primary galaxy in the Arp 82 system has a long tidal tail, a substantial bridge, and an 'ocular' waveform in its disk; all of the features are characteristic of an M51-type fly-by encounter (see \cite{kau97}). The companion galaxy is experiencing a starburst, while the primary and the tidal structures have many star-forming clumps. Considering these facts, it is somewhat surprising to find that the net Spitzer colors of the system are not unusual relative to SINGS spirals, and  do not indicate an especially high level of recent star formation. 

However, this system has other peculiarities - \cite{kau97} found that the system contains almost as much HI gas as stellar mass, an unusually high fraction. \cite{han06} fitted Starburst99 population models to GALEX and optical data, and found moderate extinctions and young ages for the clump sources. More surprisingly, evidence was found that the oldest stellar population has an age of about 2.0 Gyr in the diffuse emission. Although this result should be confirmed with near-infrared observations, and metallicity estimates should be obtained from spectral observations, it has several interesting ramifications. First of all, if the encounter has been of extended duration, most of the visible stars could have been formed in the interaction. A numerical hydrodynamical model presented in Hancock et al. suggests that the interaction could have been underway for a long time, with a first close approach about 2.0 Gyr ago. Secondly, the progenitors may have been low surface brightness galaxies with very few old stars. Given that the system is of intermediate mass, and may be forming most of its stars at about the present time, it seems to be a nearby example of ``down-sizing'' in galaxy formation. 

\section{Star Formation Processes in Interacting Galaxies: Universal or Exceptional?}

Now let us return to the questions posed at the outset. Although not a firm conclusion, it is my impression from detailed modeling and analysis of specific interacting systems over the last decade, that generally - wherever there is compression of cool gas there is SF in IGs. Moreover, the similarity (in colors and rough measures of efficiency) of SF in tidal structures and isolated disks suggests that the physical processes of SF are not very different, in these potentially very different environments. For example, they may have similar, mildly nonlinear, dependences on local mean gas density or pressure, though this point needs confirmation. Conversely, there is very little evidence from detailed studies of IGs for the existence of novel threshold behaviors or nonlinear triggering, except perhaps in the case of flows converging so hypersonically that the underlying turbulent cloud structure is completely destroyed. While we do not observe the formation of (tidal) dwarfs in isolated galaxies, their formation is plausibly explained as an extension of universal processes, e.g., the buildup of gas concentrations, gravitational instability, and more favorable environment of the persistence of marginally bound concentrations than in most galaxy disks (see Duc, this proceedings). 

At this symposium Kennicutt has presented evidence from the SINGS survey that the Kennicutt-Schmidt Law, relating SFR to a power $n$ of the gas surface density (generally $n \simeq 1.4$), extends down to kpc scales in galaxies. Gao argued that when only the densest (HCN) gas is considered, the relation is linear ($n \simeq 1.0$). If the latter conclusion is correct then the extra nonlinearity in the Kennicutt-Schmidt Law may result from a pressure or density dependence in the efficiency of dense core formation in dense clouds (see \cite{bli04}). If this is the case on small scales in dense clouds, and a nearly scale-free turbulent cascade determines the cloud dynamics  on 'intermediate' scales, {\it then the role of large-scale dynamical processes may be primarily to assemble gas concentrations and set the value of the ambient pressure.}

These tasks can be carried out by several large-scale mechanisms, including shock compression and infall and agglomeration due to gravitational instability. The recent Spitzer study of the NGC 2207/IC 2163 system by \cite{elm06} provides evidence of an agglomeration scale much larger than the scale of individual star clusters in the density waves of these galaxies. Figure 2, taken from that paper, illustrates these points. The distribution of star cluster luminosities (from HST I-band observations) is a fairly typical power-law at the more complete high luminosity (mass) end. As in other galaxies, this is probably the result of turbulent dynamics within large clouds and cloud assemblies. Clumps seen in the Spitzer $8 \mu$m image have a narrow distribution at high (summed I band) luminosity. These clumps tend to be quite uniform in each wave and are probably the result of large-scale gravitational instabilities and agglomeration. Similar phenomenology can be seen in M51 (see \cite{bas05}).

\begin{figure}
 \includegraphics[bb=-0.0cm 0.0cm 2.0cm 3.0cm, scale=2.5]{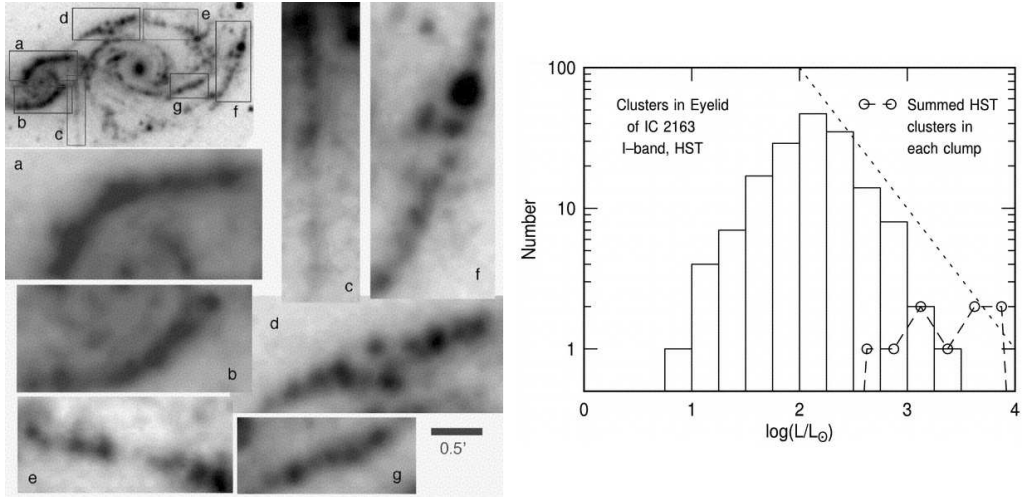}
   \caption{Left panel - Spitzer $8\mu$m logarithmic image (top left), with outlined strings of clumps enlarged below. The scale of 0.5 applies to the enlarged figures and corresponds to 5 kpc.  Right panel - luminosity distribution function in the I band of 165 star clusters observed by HST in IC 2163. The dashed line with circles is the luminosity function at I band for the sum of the clusters in each IRAC-defined ($8.0 \mu$m) star complex. Images from \cite{elm06}.
  \protect\\}
\end{figure}

While large-scale shocks can gather and compress the interstellar gas, they may also introduce further nonlinearities. For example, weak shocks may primarily affect the envelopes of molecular clouds, since they don't have the strength to overcome cloud internal pressures. The result would be increased molecule formation,  cooling, and increased star formation efficiency per unit mass of gas. Stronger shocks could compress molecular cloud cores, squeezing subcritical clumps and filaments into star formation, and probably increasing SF efficiency much more. Very strong shocks could shred and strip clouds. The densest clumps within molecular clouds would be violently compressed and heated, and so, would expand strongly afterwards. SF might well be suppressed in such cases. This limiting case might be realized collisions between disks in high speed galaxy encounters. The beautiful intergalactic shock studied recently by \cite{app06} in Stephan's Quintet might be an example. Thus, very strong shocks probably define a limit to the generalization that shock compression stimulates star formation. 

In the cases where shocks or gravitational agglomeration have stimulated SF, feedback effects may ultimately  break up the party. There is certainly evidence for this in nuclear starbursts and in dwarf irregular galaxies. In spiral density waves the rarefaction induced by large-scale divergent flow is also important. 

We can summarize the large-scale processes for gathering (or dispersing) fuel for SF with the acronym SCAF, for shock compression, agglomeration and feedback. Our understanding of how these processes accomplish large-scale stimulation in detail is still incomplete. However, almost all aspects of that understanding can be tested by observation and models in the next few years. Studies of the effects of large-scale shocks, like the modeling work described by Vazquez-Semadini at this symposium, and related observations will be particularly interesting.

\begin{acknowledgments}
I gratefully acknowledge numerous helpful and educational ``interactions'' with members of the SB\&T and Ocular collaborations. I am also grateful for support from NASA Spitzer GO grant 1263961.

\end{acknowledgments}

\end{document}